\documentclass[a4paper,11pt]{article}
\usepackage{amsfonts}
\usepackage[T1]{fontenc}
\usepackage{graphicx}
\usepackage[colorlinks]{hyperref}
\usepackage{amsmath}
\usepackage{amssymb}
\usepackage{latexsym}
\usepackage{color}
\usepackage{cite}
\usepackage{float}
\usepackage[textfont={footnotesize,sf},labelfont={color=blue,bf,sf},labelsep=endash]{caption}
\usepackage[position=top,labelfont={color=blue,bf,sf}]{subfig}
\usepackage{bm}
\usepackage{makeidx}
\usepackage{colortbl}
\usepackage{csquotes}
\usepackage[squaren]{SIunits}
\newcommand{\bra}{\begin{array}}
\newcommand{\era}{\end{array}}
\newcommand{\beq}{\begin{equation}}
\newcommand{\eeq}{\end{equation}}
\newcommand{\beqar}{\begin{eqnarray}}
\newcommand{\eeqar}{\end{eqnarray}}

\def\BC{\bb C}
\def\_\BC{\bbi C}

\def\Tr {{\rm Tr}}

\def\( {\left(}
   \def\) {\right)}
\def\[ {\left[}
\def\] {\right]}
\def\Tr {{\rm Tr}}

\newcommand{\lb}{\label}

\DeclareMathOperator{\un}{\emph{I}}
\begin{document}
\begin{titlepage}
\setcounter{page}{1}
\renewcommand{\thefootnote}{\fnsymbol{footnote}}

%\begin{flushright}
%ucd-tpg:1103.05\\
%arXiv:yymm.xxxx
%\end{flushright}

\vspace{5mm}
\begin{center}

{\Large \bf {Local quantum uncertainty of two gravitational cat states in inhomogeneous magnetic field}}

\vspace{4mm}

{\bf Rachid Hou\c{c}a$^{1,2}$\footnote{r.houca@uiz.ac.ma}
,
El Bouâzzaoui Choubabi$^{2}$\footnote{choubabi.e@ucd.ac.ma}
,
Abdellatif Kamal$^{2,3}$\footnote{abdellatif.kamal@ensam-casa.ma}
,
Abdelhadi Belouad$^{2}$\footnote{belabdelhadi@gmail.com}
and
Mohammed El Bouziani$^{2}$\footnote{elbouziani.m@ucd.ac.ma}
}

\vspace{3mm}

{$^{1}$\em Team of Theoretical Physics and High Energy, Department of Physics, Faculty of Sciences, Ibn
Zohr University}, Agadir, Morocco,\\
{\em PO Box 8106, Agadir, Morocco}

{$^{2}$\em Team of Theoretical Physics, Laboratory L.P.M.C., Department of Physics, Faculty of Sciences, Chouaib
Doukkali University, El Jadida, Morocco},\\
{\em PO Box 20, 24000 El Jadida, Morocco}

{$^{3}$\em Department of Mechanical Engineering, National Higher School of Arts and Crafts, Hassan II University, Casablanca, Morocco}

\vspace{2.5cm}

\begin{abstract}
This paper investigates the local quantum correlations (LQU), including entanglement, of two gravitational cat states subjected to an inhomogeneous magnetic field. We derived the LQU expression from the physical quantities associated with the selected system. Our findings suggest that temperature, magnetic field, and magnetic field inhomogeneity may all play a role in determining the degree of intricacy between the gravcats to some extent. Furthermore, these conclusions suggest that the thermal LQU captures a stronger quantum correlation than the entanglement. Especially true for low external magnetic field levels combined with low field inhomogeneity or high-temperature domains. Besides, we obtained the states' separability for large values of field inhomogeneity. Moreover, the correlation of the states obtained is maximal for small magnetic field values at low temperatures. Finally, we note that the state's systems become non-entangled and separable when the gap between the fundamental level and the first excited level becomes large.
\end{abstract}
\end{center}

\vspace{2cm}



\end{titlepage}

%%%%%%%%%%%%%%%%%%%%%%%%%%%%%%%%%%%%%%%%%%%%%%%%%%
\section{Introduction}
%%%%%%%%%%%%%%%%%%%%%%%%%%%%%%%%%%%%%%%%%%%%%%%%%%%%%%
Entanglement is a purely quantum correlation, having no classical analogue, between the parts of a multiparty quantum system. It is recognized as a fundamental physical resource to be exploited in many useful tasks in quantum information science \cite{Hor,Niel}. Until recent years, everyone believed that quantum correlations are closely related to quantum entanglement, and the manipulation of quantum information is generally treated in the context of entanglement and separability \cite{Hor}. However, various studies have shown that entanglement is not the only type of correlation useful for the implementation of quantum protocols, and that some separable states may also perform better than their classical counterparts \cite{Fer,Mod}. These investigations have led to the development of a new generation of quantifiers capable of detecting unclassical correlations beyond entanglement \cite{Mod,Hend,Olli,Dak}. In fact, the study of correlations in quantum systems is not limited to relating them to practical applications. Methods of quantum information theory have also been shown to be useful in the study of condensed matter systems \cite{Oste}. On the other hand, like most quantum attributes, unclassical correlations in a quantum system tend to be fragile when the system is exposed to environmental disturbances, which is inevitably the case in real situations \cite{Zurek}. In other words, all the systems we measure on are in contact with their environment. In this case, they can exchange energy with the exterior and fall into the class of dissipative systems whose evolution is the seat of irreversible processes. In addition to the dissipation of energy, which already appears in classical mechanics, the coupling of a quantum system with its environment is at the origin of the phenomenon of decoherence, which corresponds to the destruction of a coherent superposition of states during the time. This mechanism is largely responsible for the transition from the quantum world to the classical world \cite{Zurek}.

Local quantum uncertainty captures purely quantum correlations excluding their classical counterpart. This measure is quantum discord type, however with the advantage that there is no need to carry out the complicated optimization procedure over measurements. This measure is initially defined for bipartite quantum systems and a closed formula exists only for 2$\otimes$d systems. Recently, a discord-like measure has been proposed,
known as local quantum uncertainty \cite{Gir}. This measure is quantified via skew information which is achievable on a single local measurement \cite{Luo,Wig,Luoo}. This measure has a closed formula calculated for 2$\otimes$d bipartite quantum systems \cite{mez}. Later on, some authors tried to study local quantum uncertainty for orthogonally invariant class of states \cite{Sen}. This measure was also studied for quantum phase transitions \cite{Kar,Coul}. The relationship between local quantum uncertainty and quantum Fisher information under non-Markovian environment was also discussed \cite{Wu}. Recently, some authors have studied local quantum uncertainty under various decoherence models and also worked out some preliminary results for three qubits \cite{Sla1,Sla2,Sla3,Sla4}.

It is theoretically possible for a single stationary massive particle to exist in a state of combination between two spatially faraway locations, known as a Schrodinger cat state, in a quantum representation of matter. Gravitational cat (gravcat) is the name used to designate such situations for entities that gravitate. In particular, predicting the characteristics of such states is of fundamental importance in the areas of gravitational quantum physics \cite{Jour} and macroscopic quantum phenomena \cite{Karo}. In order to understand two gravitational states, much research have been carried out, the most notable of which is the study carried out by C. Anastopoulos \textit{et al}. \cite{pol} is a research facility where they have been studying the cat states produced by double-well potentials. Gravitation-induced Rabi oscillations and gravity-induced entanglement of one's own energy states describe a pair of gravcats when viewed as a two-level system each. Alternatively, they defined the non-relativistic quantum field theory in order to derive a gravitational equation of Gross-Pitaevsky for the gravcats produced in the Bose-Einstein condensation  (BEC). They used the characteristics of the two-gravcats system for BECs and its physical implications and observational possibilities by using a mathematical comparison with quantum rotors to do so. When two heavy cat states interact gravitationally, quantum correlations are generated. M. Rojas \textit{et al}. investigated the impact of a heat bath on these quantum correlations. Entanglement and quantum coherence measured by the $l_1-$norm and the competition are studied using the thermal quantum density operator \cite{rojas}.

Due to these studies, we decided to address the issue of local quantum uncertainty (LQU) in an inhomogeneous magnetic field by taking into account two gravitational cat states. Our system's Hamiltonian is simplified mathematically to identify the expression of LQU and to do this, we diagonalize our system to find the solutions to the energy spectrum, which in turn helps us find the density matrix, which is essential to finding the LQU's expressions. The current paper is organized as follows. LQU theory, which is utilized to describe this measurement explicitly, is outlined in Section \textcolor[rgb]{0.00,0.00,1.00}{2}. Section \textcolor[rgb]{0.00,0.00,1.00}{3} describes the gravitational interaction model that produces two qubits and introduces the thermal density operator owing to the thermal bath, and also determines the local quantum uncertainty as a quantifier of entanglement as a quantifier of local quantum uncertainty. Afterwards, Section \textcolor[rgb]{0.00,0.00,1.00}{4} will incorporate numerical findings and interpretations to derive conclusions and comprehend our study's aspect. In the last part, we provide a summary of the findings.
%%%%%%%%%%%%%%%%%%%%%%%%%%%%%%%%%%%%%%%%%%%%%%%%%%%%%%%%%%%%%%%%%%%%%%%%%%%%%%%%%%%%%%%%%%%%%%%%%%%
\section{Nonclassical correlation measured by local quantum uncertainty}
%%%%%%%%%%%%%%%%%%%%%%%%%%%%%%%%%%%%%%%%%%%%%%%%%%%%%%%%%%%%%%%%%%%%%%%%%%%%%%%%%%%%%%%%%%%%%%%%%%%%
A potential quantifier of non-classical correlations in multipartite systems can be considered the notion of local quantum uncertainty. This discord was mentioned to quantify the minimal quantum uncertainty generated in a quantum state due to measuring a single local observable \cite{Girolami}. Given the density matrix $\rho_{AB}\equiv\varrho$ describing a bipartite quantum state shared between two parties (say $A$ and $B$) and $\varpi_{A}^{\mu}\otimes \un_B$ denotes a local observable, with $\varpi_{A}^{\mu}$ being a Hermitian operator on $A$ with spectrum $\mu$ and $\un_B$ is the identity operator acting on the subsystem $B$. The LQU with respect to the subsystem $A$ can be defined as \cite{Girolami}
\begin{equation}\label{LQU}
\mathcal{U}(\rho_{AB}) := \min_{\varpi_{A}^{\mu} \otimes
\un_B} \mathcal{I}\Big(\rho_{AB}, \varpi_{A}^{\mu}\otimes
\un_B\Big).
\end{equation}
where the minimum is optimized over all local observables on $A$ and the good quantifier of uncertainty to an observable $\varpi^{\mu}$,
\begin{equation}\lb{skew}
\mathcal{I}(\chi,  \varpi_{A}^{\mu} \otimes
\un_B):=-\frac{1}{2}{\text{
Tr}}\left(\Big[\sqrt{\chi}, \varpi_{A}^{\mu} \otimes
\un_B\Big]^{2}\right).
\end{equation}
defines the Wigner-Yanase {\it skew information} \cite{Wig,Luo}. It serves as a measure of uncertainty of the observable $\varpi$ in the state $\chi$. For pure states $(\chi^2=\chi)$, it can be easily demonstrated that the skew information reduces to the conventional variance formula
\beq\lb{var}
\text{Var}\left(\chi, \varpi\right)=\text{Tr}\left(\chi \varpi^2\right)-\Big(\text{Tr}\big(\chi \varpi\big)\Big)^2.
\eeq
The analytical evaluation of the local quantum uncertainty is performed through a minimization procedure over the ensemble of all observable acting on the part $A$. After optimization, a closed form for qubit-qudit systems was derived in \cite{Wang}. In particular, for qubits (spin-$\frac{1}{2}$ particles), the local quantum uncertainty with respect to subsystem $A$ is given by  \cite{Girolami}
\begin{equation}\lb{lqu}
 \mathcal{U}(\rho_{AB}) = 1 - \lambda_{\max}\left\{\mathcal{W}_{AB}\right\},
\end{equation}
where $\lambda_{\max}$ denotes the maximal eigenvalue of the $3\times 3$ symmetric matrix $\mathcal{W}_{AB}$ with the entries
\begin{equation}\label{matrixomega}
 \Big(\mathcal{W}_{AB}\Big)_{lk} \equiv  {\text{
Tr}}\left\{\sqrt{\rho_{AB}}\Big(\sigma_{A}^{l}\otimes
{\un}_{B}\Big)\sqrt{\rho_{AB}}\Big(\sigma_{A}^{k}\otimes {\un}_{B}\Big)\right\}
\end{equation}
and $\sigma_{A}^{l,k} (l, k = x, y, z)$ represent the three Pauli operators of the subsystem $A$. It has been demonstrated that the LQU satisfies the full physical requirements of a measure of quantum correlations \cite{Girolami}. The local quantum uncertainty enjoys certain interesting properties. Among them, we mention its invariance under any local unitary operations. Moreover, LQU provides a reliable discord-like measure
(i.e. LQU vanishes for all states which have zero discord) and it has a geometrical significance in terms of Hellinger distance \cite{Luo,Girolami}.\\
Since the local quantum uncertainty is invariant under a local unitary transformations, so the phase factors $e^{i\alpha_{14}}$ and $e^{i\alpha_{23}}$ can be removed from the off-diagonal elements. Indeed, by means of the following local unitary transformations \cite{Kedif,Kedif1}
\begin{equation*}
|0\rangle_A=\exp\Big(-\frac{i}{2}\big(\alpha_{14}+\alpha_{23}\big)\Big)|0\rangle_A \ \ \ {\text{and}} \ \ \ |0\rangle_B=\exp\Big(-\frac{i}{2}\big(\alpha_{14}-\alpha_{23}\big)\Big)|0\rangle_B,
\end{equation*}
the density matrix $\varrho$ becomes
 \begin{equation}\label{varrhop}
\varrho\longrightarrow\varrho^{\prime} =  \left(
\begin{array}{cccc}
\rho_{11} & 0 & 0 & \vert \rho_{14} \vert \\
0 & \rho_{22} & \vert \rho_{23} \vert & 0 \\
0 & \vert \rho_{23} \vert & \rho_{33} & 0 \\
\vert \rho_{14} \vert & 0 & 0 & \rho_{44}
\end{array}
\right).
\end{equation}
The Fano–Bloch decomposition of the state $\rho$ reads
%To determine the local quantum uncertainty ${\cal{U}}(\varrho) ={\cal{U}}(\varrho^{\prime})$ defined by \eqref{lqu}, one should compute the
%eigenvalues of the matrix $\mathcal{W}_{AB}$ \eqref{matrixomega}. Explicitly, they are given by \cite{Kedif,Kedif1,Habi}
\beq
\rho={1\over4}\sum_{\alpha,\beta}R_{\alpha,\beta}\sigma_{\alpha}\otimes\sigma_{\beta}
\eeq
where the correlation matrix $R_{\alpha,\beta}$ are given by $R_{\alpha,\beta}=\Tr\left(\rho\sigma_{\alpha}\otimes\sigma_{\beta}\right)$, with $\alpha, \beta=0, 1, 2, 3$. Explicitly, they write
\begin{eqnarray}\lb{R}
R_{00} &=& \Tr(\rho)=1 \\
R_{11} &=& 2 \left(\rho _{23} +\rho _{41} \right) \\ \nonumber
R_{22} &=& 2 \left( \rho _{23} - \rho _{41} \right) \\ \nonumber
R_{33} &=& 1-2 \left(\rho _{22}+\rho _{33}\right) \\ \nonumber
R_{03} &=&1 -2 \left(\rho _{22}+\rho _{44}\right) \\ \nonumber
R_{30} &=& 1-2 \left(\rho _{33}+\rho _{44}\right) \\ \nonumber
\end{eqnarray}
Using the following relations of the Pauli matrices
\beq
\{\sigma_i,\sigma_j\}=2\delta_{ij},\quad \Tr\left(\sigma_i\sigma_j\right)=2\delta_{ij}, \quad \Tr\left(\sigma_i\sigma_j\sigma_k\sigma_l\right)=2\left(\delta_{ij}\delta_{kl}-\delta_{ik}\delta_{jl}+\delta_{il}\delta_{jk}\right)
\eeq
one shows that the matrix \eqref{matrixomega} is diagonal and the diagonal elements are
\beq
\omega_i={1\over4}\left[\sum_\beta\left(R_{0\beta}^2-\sum_kR_{k\beta}^2\right)\right]+{1\over2}\sum_kR_{i\beta}^2
\eeq
where $i, k = 1, 2, 3$ and $\beta= 0, 1, 2, 3$. They can be cast in the following closed form
\beq\lb{om}
\omega_i={1\over4}\mu^{\alpha\beta}\left(RR^t\right)_{\alpha\beta}+{1\over2}\left(RR^t\right)_{ii}
\eeq
where the summation over repeated indices is understood, the subscript $t$ stands for transposition transformation, $\mu$ is the diagonal matrix $\mu= diag \left(1,-1,-1,-1\right)$. The eigenvalues $\omega_1,\omega_1$ and $\omega_3$ \eqref{om} involve only the nonvanishing Fano–Bloch components of the square root of the density matrix $\rho$. Alternatively, they can be expanded as
\begin{eqnarray}
 \omega_1 &=&\frac{1}{4} \left(R_{00}^2-R_{33}^2+R_{03}^2-R_{30}^2+R_{11}^2-R_{22}^2\right) \\ \nonumber
 \omega_2 &=&\frac{1}{4} \left(R_{00}^2-R_{33}^2+R_{03}^2-R_{30}^2-R_{11}^2+R_{22}^2\right) \\ \nonumber
 \omega_3 &=& \frac{1}{4} \left(R_{00}^2+R_{33}^2+R_{03}^2+R_{30}^2-R_{11}^2-R_{22}^2\right)
\end{eqnarray}
where
${R}_{\mu \nu}  = {\rm Tr} \Big({\varrho^{\prime}}~\sigma_{\mu} \otimes
\sigma_{\nu} \Big)$. The non vanishing components ${R}_{\mu \nu}$ are
 \beq\lb{comp}
 \begin{array}{cc}
         R_{00} =\text{Tr}\varrho=1,~~~~~~~~~R_{03} = 1 -2 \big( \rho_{22} + \rho_{44}\big),~~~R_{30} =  1-  2 \big( \rho_{33} + \rho_{44}\big),\\
        R_{11} = 2(|\rho_{23}| + |\rho_{14}|), ~~~~~ R_{22} =  2(|\rho_{23}| - |\rho_{14}|),~~~~~R_{33} =1-  2\big(\rho_{22}+\rho_{33}\big).\\
 \end{array}
\eeq
It is must be noticed that $R_{11}\geq R_{22}$. This implies that $\omega_1>\omega_2$. Hence, from the equation \eqref{lqu}, the local quantum uncertainty quantifying the pairwise quantum correlation in the state $\varrho$ reads as
\begin{equation}\lb{lqu1}
 \mathcal{U}(\varrho) = 1 - {\max}\left\{\omega_1, \omega_3\right\}.
\end{equation}
Finally, the comparison of the eigenvalues $\omega_1$ and $\omega_3$ is a key factor in the calculation procedure of the LQU, and one needs only to find the biggest one.
%%%%%%%%%%%%%%%%%%%%%%%%%%%%%%%%%%%%%%%%%%%%%%%%%%%%
\section{Local quantum uncertainty}
%%%%%%%%%%%%%%%%%%%%%%%%%%%%%%%%%%%%%%%%%%%%%%%%%%%%%
The model consists of a set of two particles of mass $m$, each one in a one dimensional double-well potential, with local minima at $x=\pm {L\over2}$, which we will call gravitational cats. The potential is even and is such that we can associate two eigenstates for each particle describing its localization in each of the minima $|\pm\rangle$, where $x|\pm\rangle =\pm {L\over2}|\pm\rangle$. From the Landau-Lifschitz approximation \cite{pol}, they can be written in terms of the ground $|1\rangle$ and first excited states $|0\rangle$ as
\beq
|\pm\rangle={1\over\sqrt{2}}\left(|1\rangle\pm|0\rangle\right)
\eeq
The Hamiltonian $\mathcal{H}$ for this model has been analyzed in \cite{pol} and can be written as
\beq\lb{fr}
\mathcal{H}= \frac{\omega}{2}\left(\sigma_{z}\otimes\mathbb{I}+\mathbb{I}\otimes\sigma_{z}\right)-\Delta(\sigma_{x}\otimes\sigma_{x})+
\frac{B+b}{2} \left(\sigma_{z}\otimes\mathbb{I}\right)+\frac{B-b}{2}\left(\mathbb{I}\otimes\sigma_{z}\right)\
\eeq
where $\sigma_{x,z}$ are usual Pauli matrices, $\omega$ is energy difference between the ground and first excited states and furnishes the energy scale of this setup. We also have a quantity $\Delta$ that measures the intensity of the gravitational interaction between the states \cite{pol} and $B$ is the
uniform magnetic field part while $b$ indicates the degree of inhomogeneity of the magnetic field, nonuniform part of external magnetic field.
As seen below, the Hamiltonian \eqref{fr} may be represented under its matrix form in the usual computational basis $|00>$, $|01 >$, $|10 >$, $|11 >$ by
\beq\lb{1}
\mathcal{H}=\left(
\begin{array}{cccc}
 B+\omega  & 0 & 0 & -\Delta  \\
 0 & -b & -\Delta  & 0 \\
 0 & -\Delta  & b & 0 \\
 -\Delta  & 0 & 0 & -B-\omega  \\
\end{array}
\right)
\eeq
The solution of the eigenvalue equation leads to the eigenvalues, which are listed below
\begin{eqnarray}\lb{2}
\epsilon_{1,2} &=&\mp\sqrt{b^2 + \Delta^2}\\ \nonumber
\epsilon_{3,4} &=&  \mp\sqrt{(B+\omega )^2+\Delta ^2}
\end{eqnarray}
 and associated eigenvectors
\begin{eqnarray}
  |\varphi_1\rangle &=& \cos (\theta_1)|01\rangle+\sin (\theta_1)|10\rangle \\ \nonumber
  |\varphi_2\rangle &=& \cos (\theta_2)|01\rangle+\sin (\theta_2)|10\rangle \\ \nonumber
  |\varphi_3\rangle &=& \cos (\theta_3)|00\rangle+\sin (\theta_3)|11\rangle \\ \nonumber
  |\varphi_4\rangle &=& \cos (\theta_4)|00\rangle+\sin (\theta_4)|11\rangle
\end{eqnarray}
where $\theta_i$ with $i=1,2,3,4$ are defined by
\begin{eqnarray}
%  to remove numbering (before each equation)
  \theta_{1,2} &=& \arctan\left(\frac{\Delta }{b\pm\sqrt{b^2+\Delta ^2}}\right) \\ \nonumber
  \theta_{3,4} &=& \arctan\left(\frac{\Delta }{-(B+\omega )\pm\sqrt{(B+\omega )^2+\Delta ^2}}\right)
\end{eqnarray}
After knowing the spectrum of our system, it is simple to get the density matrix for our system. As a result, when the system is in thermal equilibrium, the density matrix $\rho(T)$ may be used to describe the system's state at temperature $T$. Indeed the expression of the $\rho(T)$ is given by
\beq
\rho(T)={1\over\mathbb{Z}}e^{-\beta \mathcal{H}}
\eeq
where
\beq
\mathbb{Z}=\Tr e^{-\beta \mathcal{H}}
\eeq
is the canonical ensemble partition function and $\beta={1\over k_BT}$ is the inverse thermodynamic temperature, where $k_B$ is the Boltzmann's constant, which is treated as unity in the following for the sake of simplicity.  This may be accomplished by using the spectral decomposition of the Hamiltonian \eqref{1}, which allows the thermal density matrix $\rho(T)$ to be represented as
\beq\lb{3}
\rho(T)={1\over\mathbb{Z}}\sum_{l=1}^{4}e^{-\beta E_l}|\phi_l\rangle\langle\phi_l|
\eeq
The density matrix of the system as mentioned earlier in thermal equilibrium may be expressed in the usual computational basis by substituting \eqref{2} in the equation \eqref{3}
\beq\lb{4}
\rho(T)={1\over\mathbb{Z}}\left(
\begin{array}{cccc}
 \rho_{11} & 0 & 0 & \rho_{14} \\
 0 & \rho_{22} & \rho_{23} & 0 \\
 0 & \rho_{32} & \rho_{33} & 0 \\
\rho_{41} & 0 & 0 & \rho_{44} \\
\end{array}
\right)
\eeq
\begin{eqnarray}\lb{2}
\rho_{11} &=&e^{-\beta  \epsilon _3} \cos ^2\left(\theta _3\right)+e^{-\beta  \epsilon _4} \cos ^2\left(\theta _4\right) \\ \nonumber
\rho_{14} &=& \rho_{41}=\frac{1}{2} e^{-\beta  \epsilon _3} \sin \left(2 \theta _3\right)+\frac{1}{2} e^{-\beta  \epsilon _4} \sin \left(2 \theta _4\right) \\ \nonumber
\rho_{22} &=& e^{-\beta  \epsilon _1} \cos ^2\left(\theta _1\right)+e^{-\beta  \epsilon _2} \cos ^2\left(\theta _2\right) \\ \nonumber
\rho_{23} &=&\rho_{32}=\frac{1}{2} e^{-\beta  \epsilon _1} \sin \left(2 \theta _1\right)+\frac{1}{2} e^{-\beta  \epsilon _2} \sin \left(2 \theta _2\right)\\ \nonumber
\rho_{33} &=& e^{-\beta  \epsilon _1} \sin ^2\left(\theta _1\right)+e^{-\beta  \epsilon _2} \sin ^2\left(\theta _2\right) \\ \nonumber
\rho_{44} &=& e^{-\beta  \epsilon _3} \sin ^2\left(\theta _3\right)+e^{-\beta  \text{$\epsilon $4}} \sin ^2\left(\theta _4\right)
\end{eqnarray}
whereby the partition function is defined explicitly by
\beq
\mathbb{Z}=2 \cosh \left(\beta  \sqrt{b^2+\Delta ^2}\right)+2 \cosh \left(\beta  \sqrt{(B+\omega )^2+\Delta ^2}\right)
\eeq
Because $\rho(T)$ represents a thermal state, the quantum correlations that it generates are referred to as thermal quantum correlations. The following section will quantify the quantum correlations, including entanglement, in the aforementioned two-gravcats system as a function of the system's characteristics, which include the uniform magnetic field $B$, nonuniform part of external magnetic field $b$, the excitation energy $\omega$, the intensity of the gravitational interaction $\Delta$ and  temperature $T$. The two-gravcats density matrix \eqref{4} is clearly diagonal. As a consequence, the LQU may be easily calculated using the findings mentioned earlier. To get the LQU expression, we must evaluate the eigenvalues of the matrix $\mathcal{W}_{AB}$. To beginning, one must assess the non-vanishing matrix correlation elements $R_{\mu,\nu}$ that arise during the Fano–Bloch decomposition of the density matrix $\rho(T)$. When the equations \eqref{comp} and the density matrix components \eqref{4} are combined, the result is
\begin{eqnarray}\lb{R}
R_{00} &=& \Tr(\rho)=1 \\ \nonumber
R_{11} &=& 2 \left(\rho _{23} +\rho _{41} \right) \\ \nonumber
R_{22} &=& 2 \left( \rho _{23} - \rho _{41} \right) \\ \nonumber
R_{33} &=& 1-2 \left(\rho _{22}+\rho _{33}\right) \\ \nonumber
R_{03} &=&1 -2 \left(\rho _{22}+\rho _{44}\right) \\ \nonumber
R_{30} &=& 1-2 \left(\rho _{33}+\rho _{44}\right)
\end{eqnarray}
When the equation \eqref{2} it inserted into \eqref{R}, the eigenvalues of the matrix $\mathcal{W}_{AB}$ are provided clearly by
\begin{eqnarray}
 \omega _1 &=&-2 \left(\rho _{33}^2+\left(\rho _{22}+\rho _{44}-1\right) \rho _{33}-2 \rho _{23} \rho _{41}-\rho _{22} \rho _{44}\right) \\ \nonumber
 \omega _3 &=&2 \left(\rho _{22} \left(\rho _{33}+\rho \rho _{44}-1\right)+\rho _{22}^2+\rho _{33}^2+\rho _{44}^2+\rho _{33} \rho _{44}\right)-2 \rho _{23}^2-2 \rho _{41}^2-2 \rho _{33}-2 \rho _{44}+1
\end{eqnarray}
where the elements of density matrix are defined in \eqref{2}; thus, in this instance, the LQU may be expressed in terms of $ \omega _1$ and $ \omega _3$ as follows:
\begin{equation}\lb{lqu1}
 \mathcal{U}(\varrho) = 1 - {\max}\left\{\omega_1, \omega_3\right\}.
\end{equation}
After obtaining the LQU expression, which is implicitly dependent on the inhomogeneity  $b$, the external magnetic field $B$, the energy difference between the ground and first excited states $\omega$, the intensity of the gravitational interaction $\Delta$, and the temperature $T$. As a result, we have all of the required components to investigate the behavior of the suggested system in accordance with the amounts previously mentioned. In order to demonstrate the overall performance of the proposed model, in the next section, we will devote ourselves to a numerical examination of the LQU, which will be detailed below. Next, we provided a few plots in acceptable circumstances, and we will hold different talks to bring this to a close.
%%%%%%%%%%%%%%%%%%%%%%%%%%%%%%%%%%%%
\section{Numerical results}
%%%%%%%%%%%%%%%%%%%%%%%%%%%%%%%%%%%%%%
In this part of this work, we will study the aspects of the model of two gravitational cat states numerically. Firstly, we will investigate the behavior of local quantum uncertainty as a function of temperature $T$, external magnetic field $B$, and inhomogeneity of  magnetic field $b$ by fixing the excitation energy for the value $\omega=0.05$ and also the value of intensity of the gravitation interaction for the value $\Delta=0.05$. Secondly, we will fix the temperature $T$, the magnetic field $B$ and the inhomogeneity  $b$, such that $T = B = b = 0.5$, and plot the LQU as a function of the intensity  $\Delta$ and the energy excitation $\omega$. The curves plotted are intended to make some interpretations of our proposed system. We have seen that we are dealing in units, which means that $B$, $b$, $\Delta$, $T$, and $\omega$ are all dimensionless numbers. Following that, the value $k_B=\hbar=1$ will be assumed for the sake of simplicity.
\begin{figure}[!h]
  \centering
  \includegraphics[width=6.5in]{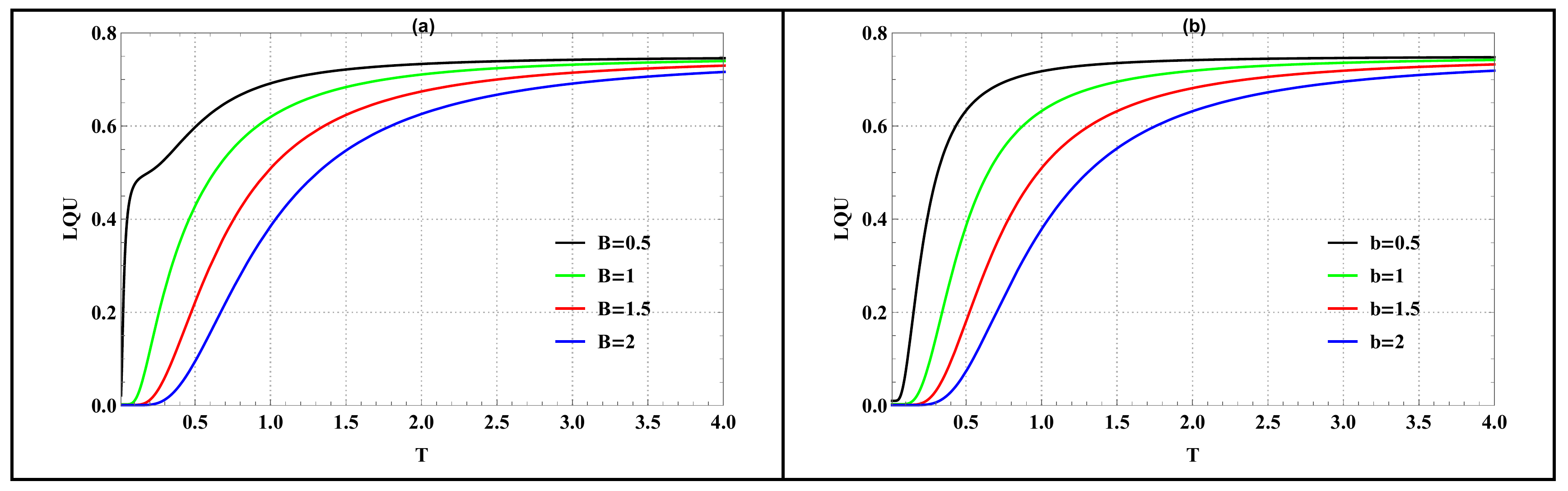}
  \caption{(Color online)(a): LQU versus $T$ for various values of $B$. (b): for different values of $b$ for fixed values of $\omega$ and $\Delta$, such that $\omega=0.05$ and $\Delta=0.05$.}\label{f1}
\end{figure}

In figure \eqref{f1} the common remark between the two figures  (\ref{f1}.a) and (\ref{f1}.b) at low temperature that the LQU is zero. which means that the states of the systems are not correlated, and therefore, they are separable. However, at high temperatures, the LQU tends towards a fixed value of less than one even when the magnetic fields and the inhomogeneity increase; most of the gravcats are correlated. It is about $70\%$ and that the rest, which represents $30\%$ of the gravcats, are separable. On the other hand, when $T$ starts to increase, the LQU increases to a constant value equal to $0.7$. Still, the influence of the magnetic field $B$ and the inhomogeneity $b$ on the LQU disappeared for high values of the temperature $T$, which is easy to see in the two figures above.  Furthermore, for large values of $B$ and $b$ in low temperature, there is the appearance of a temperature interval where  LQU remains zero, and the interval width increases according to the values of $B$ and $b$. To explain, we deduce that the temperature $T$,  the magnetic field $B$ and the inhomogeneity $b$ can be responsible for more or less of the intricacy between the gravcats.

\begin{figure}[!h]
  \centering
  \includegraphics[width=6.5in]{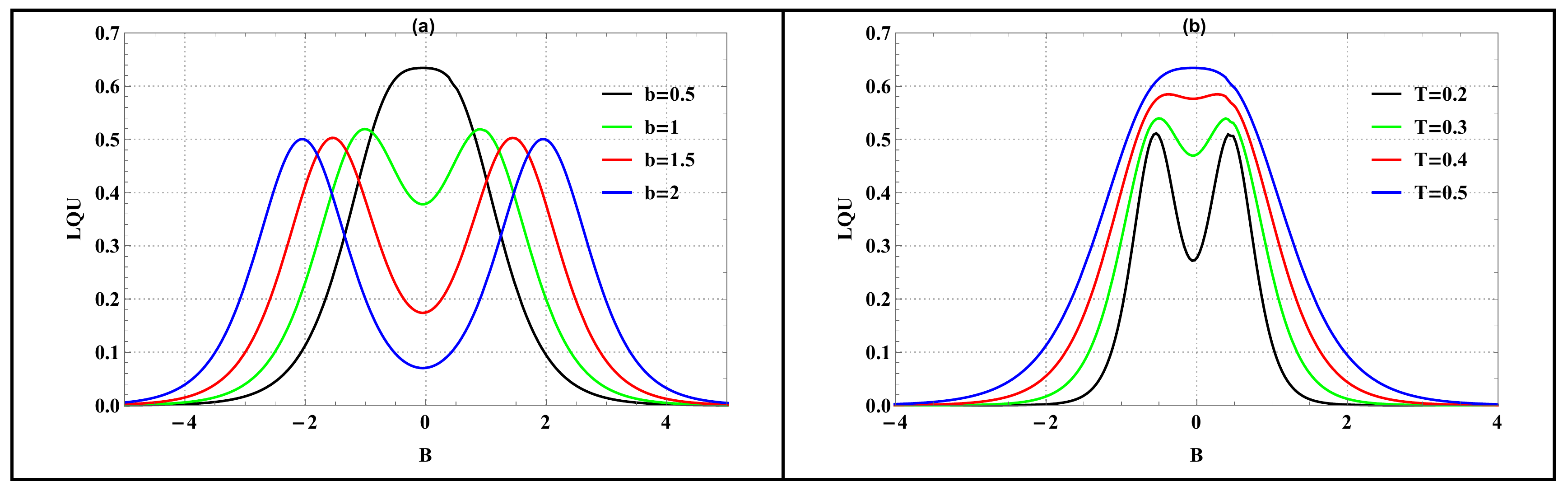}
  \caption{(Color online)(a): LQU versus $B$ for various values of $b$. (b): for different values of $T$ for fixed values of $\omega$ and $\Delta$, such that $\omega=0.05$ and $\Delta=0.05$.}\label{f2}
\end{figure}
From the figure \ref{f2}, the common observation between the two figures is that the thermal quantum correlations are symmetrical
zero magnetic fields $B=0$, and the LQU tends to zero for the high external magnetic field $B$ values. Moreover, for the small nonuniform magnetic field $b$ and high temperature $T$ when $B$ tends to zero, can give rise to higher quantum correlations its shown respectively, in the black curve in (\ref{f2}.a) and blue curve in (\ref{f2}.b). Then we can conclude that the thermal LQU captures a quantum correlation exceeding the entanglement, particularly for weak external magnetic field values together with the weak values of the inhomogeneity of the magnetic field $b$ or high-temperature regime. In addition, due to the maximization process \eqref{lqu1}, LQU exhibits a sudden change with a double peak structure. This latter characteristic incorporates the LQU, unusually when the two-gravcats system is subjected only to a nonuniform magnetic field $b$.
\begin{figure}[!h]
  \centering
  \includegraphics[width=6.5in]{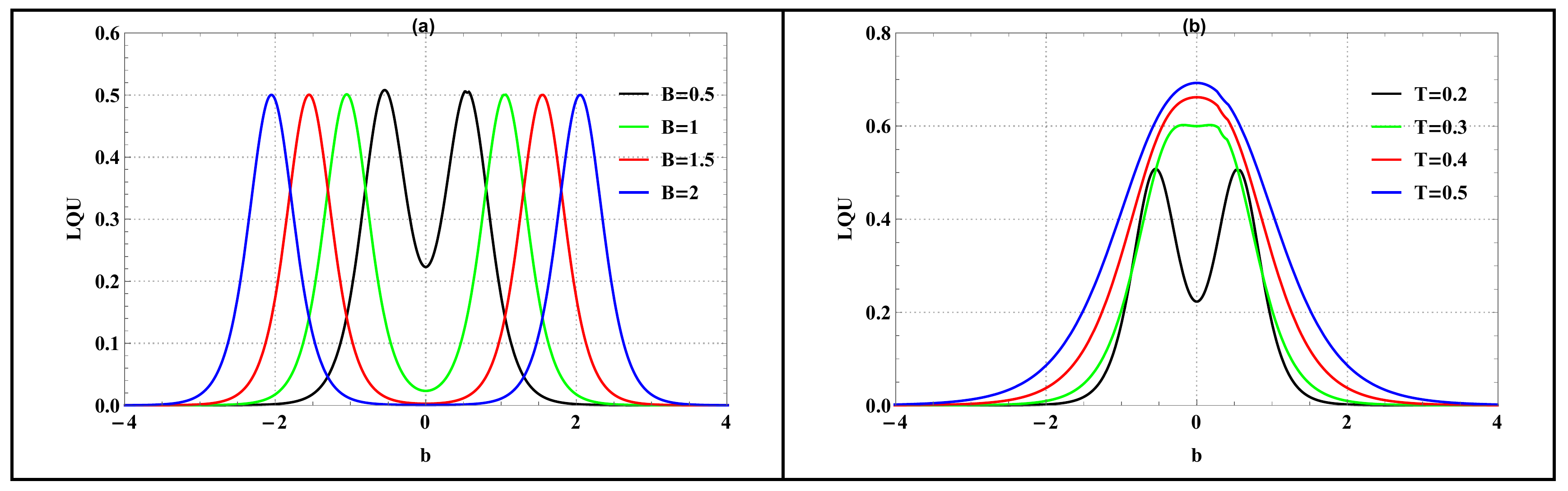}
  \caption{(Color online)(a): LQU versus $b$ for various values of $B$. (b): for different values of $T$ for fixed values of $\omega$ and $\Delta$, such that $\omega=0.05$ and $\Delta=0.05$.}\label{f3}
\end{figure}

Figure \eqref{f3}, the LQU is always symmetrical concerning $b = 0$. For large values of $b$, the LQU quickly tends towards zero when the magnetic field $B$ becomes weak. Its shown in the black curve in figure (\ref{f3}.a), and likewise, for low temperature, see the black curve in figure (\ref{f3}.b). Moreover, the figure (\ref{f3}.a) presents two peaks having the same amplitude. The separation between the two peaks increases when the magnetic field increases and vice versa. For $B=0$, we will have an overlap of two peaks which gives a salt maximum peak centered in the middle. In figure (\ref{f3}.b), always two peaks move away from each other when the temperature decreases and vice versa, the overlap occurs in this case at low temperature. In conclusion, the separability of the states can be obtained for large values of the inhomogeneity of the field $b$; accordingly, the correlation of the states will be maximal for small values of the magnetic field $ B $ at low $T$.
\begin{figure}[!h]
  \centering
  \includegraphics[width=6.5in]{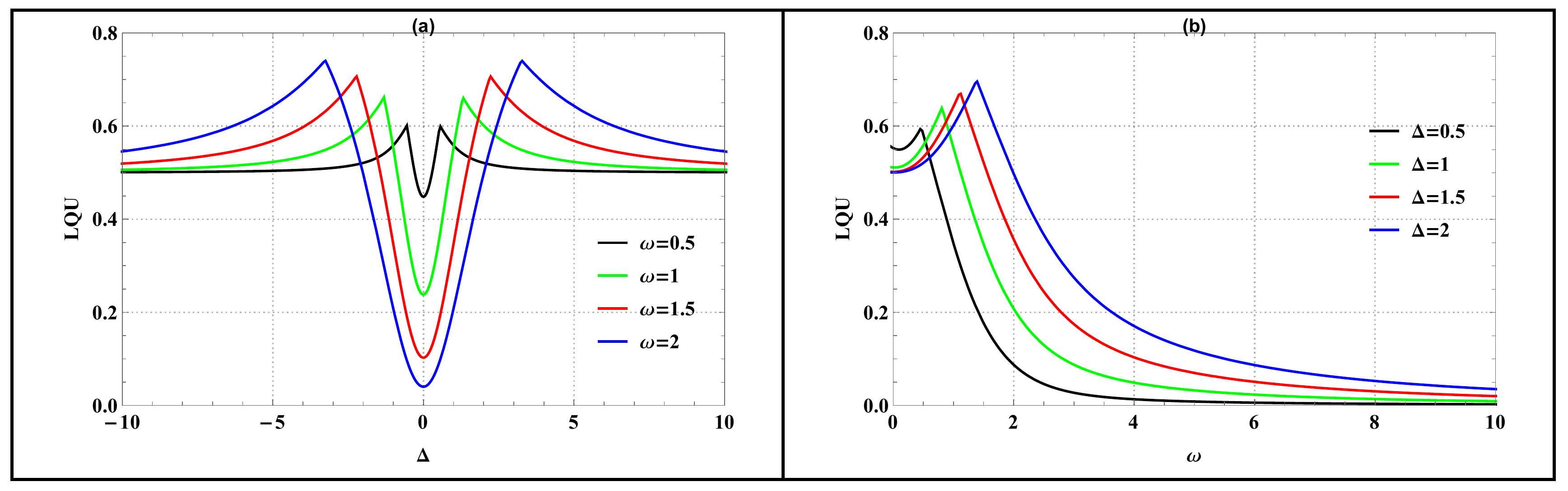}
  \caption{(Color online)(a): LQU versus $T$ for various  values of $\Delta$. (b): for different  values of $\omega$  for fixed values of $B$, $b$ and $T$, such that $B=b=T=0.5$ .}\label{f4}
\end{figure}

Figure (\ref{f4}.a) shows that the symmetry is always maintained concerning $\Delta =0$. For high values of $\Delta$, LQU tends to a constant value equal to $0.5$, regardless of how much the excitation energy $\omega$ rises. As $\Delta =0$,  LQU decreases towards zero when the difference between the ground state and the first excited state rises. However, when $\Delta$ increases, the LQU increases until it reaches a maximum value, after which it declines towards zero. Figure (\ref{f4}.b) illustrates that for small values of $\omega$, the LQU increases until it reaches a maximum that depends on the gravitational intensity $\Delta$; in this instance, the states are highly correlated, afterwards, the LQU eventually decreases to zero for high values of $\omega$, or the states become completely separable.
%%%%%%%%%%%%%%%%%%%%%%%%%%%%%%%%%%%%%%%%%%%%%%%%%%%%%%%%%%%
\section{Conclusion}
%%%%%%%%%%%%%%%%%%%%%%%%%%%%%%%%%%%%%%%%%%%%%%%%%%%%%%%%%%%%%
Local quantum correlations, including entanglement, of two gravitational cat states, are investigated when an inhomogeneous magnetic field is applied. The Hamiltonian model is given, and through mathematical calculations, the eigenstates entanglement have been determined, and the thermal state at a finite temperature is explicitly derived. Then, the LQU expression has been obtained in terms of the magnetic field $B$, inhomogeneity $b$, temperature $T$, excitation energy $\omega$, and the intensity of gravity $\Delta$, the behavior of the thermal quantum correlations for our study have been investigated numerically. In this paper, we have concluded that the temperature $T$, the magnetic field $B$, and the inhomogeneity $b$ may all play a role in determining the degree of intricacy between the gravcats to a greater or lesser extent. Moreover, it is possible to infer from these results that the thermal LQU captures a more significant quantum correlation than the entanglement, which is particularly true for low external magnetic field levels combined with low values of inhomogeneity of the magnetic field $b$ or for high-temperature domains. Again, the separability of the states can be obtained for large values of the inhomogeneity of the field $b$; accordingly, the correlation of the states will have been obtained maximal for small values of the magnetic field $ B $ at low $T$. At long last, it has been found that when the gap between the fundamental level and the first excited level grows in size, the LQU tends to become zero, where the state's systems become non-entangled.
%%%%%%%%%%%%%%%%%%%%%%%%%%%%%%%%%%%%%%%%%

\section*{Acknowledgment}

%%%%%%%%%%%%%%%%%%%%%%%%%%%%%%%%%%%%%%%

Our sincere gratitude goes to Mohamed Monkad, head of the Laboratory for Physics of Condensed Matter (LPMC) in the Faculty of Sciences at Choua\"ib Doukkali University, for his essential assistance.

%%%%%%%%%%%%%%%%%%%%%%%%%%%%%%%%%%%%%%%%%%%%%%%%%%%%%%%%%%%%%%%%%%%%%%%%%%

\end{document}